\documentclass[12pt]{article}
\usepackage{amsmath,amsfonts,amssymb,bm}
\usepackage{graphicx}

\title{A Perturbative Method for Nonequilibrium Steady State of Open Quantum Systems}

\author{Tatsuro Yuge
\\
\normalsize Department of Physics, Osaka University, 1-1, Toyonaka, Osaka, 560-0043, Japan
\bigskip\\
Ayumu Sugita
\\
\normalsize Department of Applied Physics, Osaka City University, Sugimoto, Osaka, 558-8585, Japan
}

\begin{document}

\maketitle

\begin{abstract}
We develop a method of calculating the nonequilibrium steady state (NESS) of an open quantum system 
that is weakly coupled to reservoirs in different equilibrium states. 
We describe the system using a Redfield-type quantum master equation (QME). 
We decompose the Redfield QME into a Lindblad-type QME and the remaining part $\mathcal{R}$. 
Regarding the steady state of the Lindblad QME as the unperturbed solution, 
we perform a perturbative calculation with respect to $\mathcal{R}$ to obtain the NESS of the Redfield QME. 
The NESS thus determined is exact up to the first order in the system-reservoir coupling strength (pump/loss rate), 
which is the same as the order of validity of the QME. 
An advantage of the proposed method in numerical computation is 
its applicability to systems larger than those in methods of directly solving the original Redfield QME. 
We apply the method to a noninteracting fermion system 
to obtain an analytical expression of the NESS density matrix. 
We also numerically demonstrate the method in a nonequilibrium quantum spin chain. 
\end{abstract}

\section{Introduction}

Establishing the statistical mechanics of the nonequilibrium steady state (NESS) is 
one of the most challenging problems in physics. 
In analogy to equilibrium statistical mechanics, 
a possible and desirable answer to this problem would be an ensemble theory 
with which we could write down the density matrix (or distribution function) for a NESS. 
To proceed in this direction, it is important to extract the characteristics of density matrices for NESSs 
by constructing and analyzing NESSs in various systems.

To realize a NESS in a concrete system, we should put the system in contact with heat baths or reservoirs 
that absorb the energy dissipated from the system. 
We thus have to treat open systems to analyze NESS. 
In quantum systems, one of the theoretical frameworks widely used for open systems 
is the quantum master equation (QME) \cite{BreuerPetruccione}, 
an equation of motion for the density matrix of the system. 
In fact, the QME is used in various fields of physics: e.g., 
quantum optics \cite{BreuerPetruccione,Carmichael}, nuclear magnetic resonance \cite{Slichter}, 
electron transfer in chemical physics and biophysics \cite{Pollard_etal,Kondov_etal_2}, 
heat transport \cite{Saito_etal_1,Saito_etal_2,Saito,Wichterich_etal,Mejia-MonasterioWichetrich,Michel_etal,
Yan_etal,ProsenZnidaric_1,WuBerciu}, 
electronic transport in mesoscopic conductors \cite{GurvitzPrager,Fischetti,Harbola_etal,Gudmundsson_etal,Yuge_etal_1}, 
spin transport \cite{ThingnaWang}, 
and nonequilibrium thermodynamics and statistical physics 
\cite{SpohnLebowitz,KuboTodaHashitsume,EspositoHarbolaMukamel_RMP,ChetriteMallick,Yuge_etal_2}. 
Therefore, the QME is a reliable approach to investigating NESS in various systems (c.f. Ref.~\cite{Sugita}). 
In some solvable models, analytical expressions of the NESS of the QME are obtained 
\cite{Prosen_1,ProsenZnidaric_2,Prosen_2,Prosen_3,Karevski_etal,Prosen_4}. 

Apart from solvable models, however, there appears a difficulty in analyzing the NESS of the QME. 
This is because one should treat a huge number of components in the QME; 
the number of elements of a density matrix (mixed state) is the square of that of a wave function (pure state). 
This restricts available system size to be relatively small when one employs direct methods 
(direct diagonalization or time integration) in solving the QME, even though efficient numerical methods are developed 
\cite{Pollard_etal,Kondov_etal_1}. 

A method of treating larger systems in the framework of the QME 
is the stochastic unraveling of QME (also known as quantum jump, quantum trajectory, 
and Monte Carlo wave function method) \cite{BreuerPetruccione,Carmichael,PlenioKnight,Dalibard_etal,Dum_etal,
Gardiner_etal,GisinPercival,Strunz_etal,Breuer_etal,Kleinekathofer_etal,GaspardNagaoka_2,Kondov_etal_2,
Mejia-MonasterioWichetrich,Michel_etal}. 
In this method, since one treats a wave function instead of a density matrix, 
one can investigate systems larger than those in the direct methods. 
This method, however, requires a Monte Carlo calculation; 
i.e., one must take an average of many times of iterative computations. 
There are also other methods.
In Ref.~\cite{ProsenZnidaric_1}, 
a density matrix renormalization group method with a matrix product operator ansatz is proposed. 
This method assumes the (rather phenomenological) local action of the dissipators associated with the baths. 
In Ref.~\cite{Wu}, a many-particle Green function method is proposed. 
This requires the approximation of weak internal coupling or small system size. 

In this paper, we propose another method of calculating the NESS of the QME. 
We use a perturbation theory with respect to the coupling strength (pump/loss rate) 
between the system and the reservoirs. 
Our method assumes only the weak system-reservoir coupling, which is the same assumption as that of the QME. 
In this method, the unperturbed part of the NESS has only diagonal elements in the energy eigenstate basis. 
Thus, the number of elements of the unperturbed part is on the same order as that of the wave function. 
We can therefore treat systems as large as those in a stochastic unraveling method. 
Moreover, since this method does not require Monte Carlo calculation, 
it is expected to be faster than the unraveling method. 
We also note that the Markovian approximation use in the QME holds exactly for the steady-state solution, 
since the approximation is good when the system changes slowly \cite{KuboTodaHashitsume}.

In the present paper, we describe the method in a setup of a system that is weakly coupled to two reservoirs. 
In the next section, we explain the setup and QME. 
In Sect.~\ref{section:main}, we give the main result, i.e., the perturbative solution for the NESS of the QME. 
We also show an advantage of our method in numerical computation. 
Furthermore, we explain how we calculate currents in this method. 
In Sect.~\ref{section:example}, we have two examples. 
One is a noninteracting fermion system, where we derive an analytical expression of the NESS by the method. 
The other is a numerical computation of the NESS in a quantum spin chain, 
where we demonstrate the validity of the method. 
We devote Sect.~\ref{section:conclusion} to the concluding remarks.

\section{Setup}\label{section:setup}

\begin{figure}[t]
\begin{center}
\includegraphics[width=0.6\linewidth]{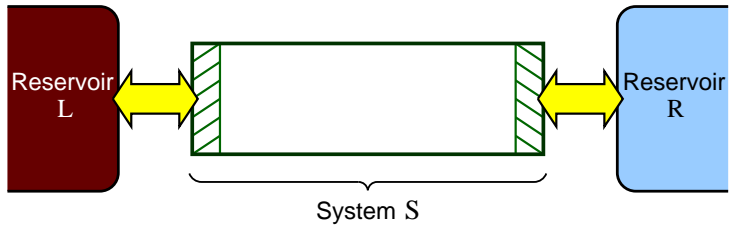}
\end{center}
\caption{
(Color online) Schematic diagram of the setup. The system S is connected to two reservoirs, L and R. 
The reservoir $b$ ($b={\rm L,R}$) is in an individual equilibrium state. 
The hatched regions in S represent the interaction regions with the reservoirs.  
}
\label{fig:schematic}
\end{figure}

We consider transport phenomena (energy transport, particle transport, {\it etc.}) in the quantum system S. 
S is in contact with two reservoirs (heat baths), L and R (see Fig.~\ref{fig:schematic}), 
and is thus an open system. 
We assume that the dimension of the Hilbert space $\mathsf{H}$ associated with S is finite 
($\dim\mathsf{H} < \infty$). 
Each reservoir is sufficiently large compared with S and is in an equilibrium state characterized by 
its own inverse temperature $\beta_b$, chemical potential $\mu_b$, and so on ($b={\rm L,R}$). 
We also assume that the coupling between the system S and each reservoir is weak. 
The Hamiltonian of the total system (L+S+R) reads 
\begin{align}
\hat{H}_{\rm tot} = \hat{H}_{\rm S} + \sum_{b={\rm L,R}} \bigl( \hat{H}_b + u \hat{H}_{{\rm S}b} \bigr). 
\label{Htot}
\end{align}
The first term $\hat{H}_{\rm S}$ is the Hamiltonian of S, which may be degenerate. 
We denote the eigenenergy as $E_i$ and the corresponding eigenstate as $|E_i,m_i\rangle$, 
where $m_i$ is the label for distinguishing the degenerate states. 
The second term $\hat{H}_b$ is the Hamiltonian of the reservoir $b$. 
The third term $\hat{H}_{{\rm S}b}$ is the system-reservoir coupling Hamiltonian. 
The small prefactor $u$ represents the weak system-reservoir coupling assumption. 

\subsection{Quantum master equation}

Here, we explain the quantum master equation (QME) to fix the notation in the present paper. 
The starting point is the Liouville-von Neumann equation of the total system: 
\begin{align}
\frac{d\hat{\rho}_{\rm tot}(t)}{dt} 
= \frac{1}{i\hbar} [\hat{H}_{\rm tot} , \hat{\rho}_{\rm tot}(t)], 
\label{Liouville-vonNeumann}
\end{align}
where $\hat{\rho}_{\rm tot}$ represents the density matrix of the total system. 
We can derive the QME, the equation of motion of the system S, 
by applying the Born-Markov approximation \cite{BreuerPetruccione} to Eq.~(\ref{Liouville-vonNeumann}). 
The QME in the Schr\"odinger picture reads 
\begin{align}
\frac{d\hat{\rho}(t)}{dt} = \mathcal{L} \hat{\rho}(t). 
\label{QME}
\end{align}
Here, $\hat{\rho} = {\rm Tr_L Tr_R} \hat{\rho}_{\rm tot}$ is the reduced density matrix of S, 
and ${\rm Tr}_b$ is the trace over the reservoir $b$. 
The superoperator (QME generator) $\mathcal{L}$ is given by 
\begin{align}
\mathcal{L} \equiv \mathcal{L}_0 + v \sum_b \mathcal{L}_b, 
\label{generator1}
\end{align}
where $v \equiv u^2$ is a parameter that controls the pump/loss rate due to the coupling to the reservoirs, 
$\mathcal{L}_0 \hat{\rho} \equiv [\hat{H}_{\rm S}, \hat{\rho}]/i\hbar$ describes the unitary time evolution, and 
\begin{align}
\mathcal{L}_b \hat{\rho} \equiv 
- \frac{1}{\hbar^2} \int_0^\infty dt' {\rm Tr}_b \Bigl[ \hat{H}_{{\rm S}b} , \bigl[ \breve{H}_{{\rm S}b}(-t') , 
\hat{\rho} \otimes \hat{\rho}_b \bigr] \Bigr] 
\label{dissipator}
\end{align}
is the dissipative part associated with the reservoir $b$. 
Here, $\breve{O}(\tau) = \hat{U}^\dag(\tau) \hat{O} \hat{U}(\tau)$ 
with $\hat{U}(\tau) \equiv \exp \bigl\{ - \bigl( \hat{H}_{\rm S} 
+ \sum_b  \hat{H}_b \bigr) \tau /i\hbar\bigr\}$ is an operator in the interaction picture, 
and $\hat{\rho}_b$ is the equilibrium state of the reservoir $b$. 
The explicit form of $\hat{\rho}_b$ depends on situations; 
the canonical ensemble 
$\hat{\rho}_b = \exp \bigl( - \beta_b \hat{H}_b \bigr) \big/ {\rm Tr}_b \exp \bigl( - \beta_b \hat{H}_b \bigr)$ 
or the grand canonical ensemble 
$\hat{\rho}_b = \exp \bigl[ - \beta_b (\hat{H}_b - \mu_b \hat{N}_b ) \bigr] 
\big/ {\rm Tr}_b \exp \bigl[ - \beta_b (\hat{H}_b - \mu_b \hat{N}_b ) \bigr]$ is usually used. 
In deriving these equations, we assumed ${\rm Tr}_b [\hat{\rho}_b \hat{H}_{{\rm S}b}] = 0$.
Equation (\ref{QME}) with Eq.~(\ref{generator1}) is called Redfield QME. 
We note that the QME is valid up to first order in $v$ ($=u^2$) 
since the Born approximation is a second-order approximation with respect to $u$.

We now rewrite the QME into a more tractable form. 
To this end, we first assume that the system-reservoir coupling Hamiltonian has the form 
\begin{align}
\hat{H}_{{\rm S}b} = \sum_\lambda \hat{X}_{b,\lambda} \otimes \hat{Y}_{b,\lambda}, 
\label{interaction}
\end{align}
where $\hat{X}_{b,\lambda}$ is a self-adjoint operator of S that is defined 
locally in the regions (hatched in Fig.~\ref{fig:schematic}) near the reservoir $b$, 
and $\hat{Y}_{b,\lambda}$ is a self-adjoint operator of the reservoir $b$. 
We next decompose $\hat{X}_{b,\lambda}$ into eigenoperators of $\hat{H}_{\rm S}$ \cite{BreuerPetruccione}: 
\begin{align}
\hat{X}_{b,\lambda} &= \sum_\omega \hat{X}_{b,\lambda}^{(\omega)}, 
\label{eigenOpDecomposition}
\\
\hat{X}_{b,\lambda}^{(\omega)} &\equiv \sum_i \hat{\Pi}(E_i) \hat{X}_{b,\lambda} \hat{\Pi}(E_i + \hbar\omega), 
\end{align}
where $\hbar\omega$ runs over the eigenenergy differences, 
and $\hat{\Pi}(E_i) = \sum_{m_i} |E_i, m_i\rangle \langle E_i, m_i|$ is the projection operator onto the eigenspace 
with the eigenenergy $E_i$. 
Substituting Eqs.~(\ref{interaction}) and (\ref{eigenOpDecomposition}) into Eq.~(\ref{dissipator}), we have 
\begin{align}
\mathcal{L}_b \hat{\rho} 
&= \frac{1}{\hbar^2} \sum_{\lambda,\nu} \sum_\omega \Xi_{\lambda\nu}^b (\omega) 
\Bigl\{ 
\hat{X}_{b,\nu}^{(\omega)\dag} \hat{\rho} \hat{X}_{b,\lambda} 
- \hat{X}_{b,\lambda} \hat{X}_{b,\nu}^{(\omega)\dag} \hat{\rho} 
\Bigr\} 
+ {\rm h.c.}
\nonumber\\
&= \frac{1}{\hbar^2} \sum_{\lambda,\nu} \sum_{\omega,\omega'} \Xi_{\lambda\nu}^b (\omega) 
\Bigl\{ 
\hat{X}_{b,\nu}^{(\omega)\dag} \hat{\rho} \hat{X}_{b,\lambda}^{(\omega')} 
- \hat{X}_{b,\lambda}^{(\omega')} \hat{X}_{b,\nu}^{(\omega)\dag} \hat{\rho} 
\Bigr\} 
+ {\rm h.c.}, 
\label{dissipator2}
\end{align}
where $\Xi_{\lambda\nu}^b (\omega)$ is the Fourier-Laplace transform of the reservoir correlation function: 
\begin{align}
\Xi_{\lambda\nu}^b (\omega) 
\equiv \int_0^\infty dt e^{-i\omega t} {\rm Tr}_b \bigl[ \breve{Y}_{b,\lambda}(t) \hat{Y}_{b,\nu} \hat{\rho}_b \bigr].
\end{align}
In Eq.~(\ref{dissipator2}), we used $\hat{X}_{b,\lambda}^\dag = \hat{X}_{b,\lambda}$ 
and $\hat{X}_{b,\lambda}^{(\omega)\dag} = \hat{X}_{b,\lambda}^{(-\omega)}$.

\subsection{Decomposition of QME}

From Eq.~(\ref{dissipator2}), 
we can decompose $\mathcal{L}_b$ into the part with $\omega=\omega'$ and the remaining one: 
$\mathcal{L}_b = \mathcal{L}_b^{\rm SA} + \mathcal{R}_b$, where 
\begin{align}
\mathcal{L}_b^{\rm SA} \hat{\rho} 
&\equiv \frac{1}{\hbar^2} \sum_{\lambda,\nu} \sum_{\omega} \Xi_{\lambda\nu}^b (\omega) 
\Bigl\{ 
\hat{X}_{b,\nu}^{(\omega)\dag} \hat{\rho} \hat{X}_{b,\lambda}^{(\omega)} 
- \hat{X}_{b,\lambda}^{(\omega)} \hat{X}_{b,\nu}^{(\omega)\dag} \hat{\rho} 
\Bigr\} 
+ {\rm h.c.} 
\nonumber\\
&= \frac{1}{i\hbar} [\hat{H}_{\rm LS}^b, \hat{\rho}] 
+ \frac{1}{2\hbar^2} \sum_{\lambda,\nu} \sum_\omega \Phi_{\lambda\nu}^b(\omega) 
\Bigl\{ 
2 \hat{X}_{b,\nu}^{(\omega)\dag} \hat{\rho} \hat{X}_{b,\lambda}^{(\omega)} 
- \hat{X}_{b,\lambda}^{(\omega)} \hat{X}_{b,\nu}^{(\omega)\dag} \hat{\rho} 
- \hat{\rho} \hat{X}_{b,\lambda}^{(\omega)} \hat{X}_{b,\nu}^{(\omega)\dag} 
\Bigr\}. 
\end{align}
Here, $\hat{H}_{\rm LS}^b \equiv \sum_{\lambda,\nu} \sum_\omega 
\Psi_{\lambda\nu}^b(\omega) \hat{X}_{b,\lambda}^{(\omega)} \hat{X}_{b,\nu}^{(\omega)\dag}/2\hbar$ 
is a Lamb shift Hamiltonian, 
$\Phi_{\lambda\nu}^b(\omega) \equiv \Xi_{\lambda\nu}^b(\omega) + \overline{\Xi_{\nu\lambda}^b}(\omega)$, 
$\Psi_{\lambda\nu}^b(\omega) \equiv -i[ \Xi_{\lambda\nu}^b(\omega) - \overline{\Xi_{\nu\lambda}^b}(\omega) ]$, 
and the overlines stand for the operation of taking the complex conjugate. 
Since $\Phi_{\lambda\nu}^b(\omega)$ is positive definite, $\mathcal{L}_b^{\rm SA}$ gives a Lindblad-type QME. 
That is, the $\mathcal{R}_b$-omitted superoperator 
$\mathcal{L}_{\rm SA} = \mathcal{L}_0 + v \sum_b \mathcal{L}_b^{\rm SA}$ 
is a generator of a completely positive dynamical semigroup \cite{Gorini_etal,Lindblad}. 
The approximation of omitting the $\mathcal{R}_b$ terms is called secular approximation (SA) 
or rotating wave approximation \cite{BreuerPetruccione}. 
The SA is an approximation that extracts a slow part of the dynamics; 
one can obtain $\mathcal{L}_{\rm SA}$ by averaging $\mathcal{L}$ in time in the interaction picture or 
by omitting the fast oscillating terms in $\mathcal{L}$ in the interaction picture. 
However, it is known that the internal current vanishes in the NESS of the SA-QME \cite{Wichterich_etal}. 
Therefore, the SA-QME is not appropriate for analyzing the steady state itself (however, note Ref.~\cite{note1}), 
and one often uses any of the original Redfield QME, alternatively approximated Lindblad QME \cite{Wichterich_etal}, 
or axiomatic Lindblad QME \cite{Gorini_etal,Lindblad} for nonequilibrium situations. 

In the present study, we use the Redfield QME. 
From the above argument, we have a decomposition of the QME generator: 
\begin{align}
\mathcal{L} = \mathcal{L}_{\rm SA} + v \mathcal{R}, 
\label{generator2}
\end{align}
with $\mathcal{R} \equiv \sum_b\mathcal{R}_b$. 
In the next section, we develop a perturbative method of calculating the NESS of the Redfield QME (\ref{QME}), 
where we regard $\mathcal{L}_{\rm SA}$ and $v \mathcal{R}$ respectively as the unperturbed and perturbation parts.

\subsection{Liouville space and projection superoperator}

To summarize the points so far, 
we have introduced the Redfield QME (\ref{QME}) that describes the dynamics of the open quantum system S. 
The QME generator $\mathcal{L}$ is written as Eq.~(\ref{generator1}) or (\ref{generator2}). 
Before going to the main result in the next section, here, we make two more preliminaries. 

One is the Liouville space $\mathsf{L}$, which is the set of all the linear operators on $\mathsf{H}$. 
Since the dimension of $\mathsf{H}$ is finite, any $\hat{A}$ in $\mathsf{L}$ is a trace class operator.
We can define the Hilbert-Schmidt inner product in $\mathsf{L}$ as 
${\rm Tr}_{\rm S}(\hat{A}_1^\dag \hat{A}_2)$ for any $\hat{A}_1, \hat{A}_2 \in \mathsf{L}$, 
where ${\rm Tr}_{\rm S}$ is the trace in $\mathsf{H}$. 
With this inner product, $\mathsf{L}$ is a Hilbert space. 
We refer to any linear operator on $\mathsf{L}$ as a superoperator 
(which includes $\mathcal{L}, \mathcal{L}_{\rm SA}$, and so on). 
We note that the dimension of $\mathsf{L}$ is given as $\dim\mathsf{L} = \dim\mathsf{H}^2$. 
We define the adjoint $\mathcal{O}^\dag$ of a superoperator $\mathcal{O}$ such that  
${\rm Tr}_{\rm S} [(\mathcal{O}^\dag \hat{A}_1)^\dag \hat{A}_2] 
= {\rm Tr}_{\rm S} (\hat{A}_1^\dag \mathcal{O} \hat{A}_2)$ 
holds for any $\hat{A}_1, \hat{A}_2 \in \mathsf{L}$.

The other is the following projection superoperator $\mathcal{P}$, which is defined by 
\begin{align}
\mathcal{P} |E_i, m_i\rangle \langle E_j, m_j| = \left\{
\begin{array}{ll}
|E_i, m_i\rangle \langle E_j, m_j| & (E_i = E_j) 
\\
0 & (E_i \neq E_j). 
\end{array}
\right.
\end{align}
In the matrix representation of any $\hat{A} \in \mathsf{L}$ 
in the basis of the energy eigenstates of $\hat{H}_{\rm S}$, 
$\mathcal{P}$ extracts only the matrix elements constructed from the eigenstates with the same eigenenergies. 
We also define the projection superoperator $\mathcal{Q} \equiv 1-\mathcal{P}$. 
By using these superoperators, we decompose $\mathsf{L}$ into the subspace 
$\mathsf{P} \equiv \{ \hat{A} \in \mathsf{L} | ~ \mathcal{P}\hat{A} = \hat{A} \}$
and its orthogonal complement $\mathsf{Q}$. 
We note that $\dim\mathsf{P}$ is on the same order as $\dim\mathsf{H}$ ($\ll \dim\mathsf{L}$), 
particularly, $\dim\mathsf{P} = \dim\mathsf{H}$ if eigenenergies of $\hat{H}_{\rm S}$ are not degenerate. 

By using the above projection superoperators, we can show the following relations: 
\begin{align}
\mathcal{P} \mathcal{L}_0 &= \mathcal{L}_0 \mathcal{P} = 0, 
\\ 
\mathcal{P} \mathcal{L}_b^{\rm SA} \mathcal{Q} &= \mathcal{Q} \mathcal{L}_b^{\rm SA} \mathcal{P} 
= \mathcal{P} \mathcal{L}_{\rm SA} \mathcal{Q} = \mathcal{Q} \mathcal{L}_{\rm SA} \mathcal{P} = 0, 
\label{SA-L}
\\
\mathcal{P} \mathcal{R}_b \mathcal{P} &= \mathcal{P} \mathcal{R} \mathcal{P} = 0. 
\label{PRP}
\end{align}
Equation~(\ref{SA-L}) leads to $\mathcal{L}_{\rm SA} = \mathcal{P} \mathcal{L}_{\rm SA} \mathcal{P} 
+ \mathcal{Q} \mathcal{L}_{\rm SA} \mathcal{Q}$, 
which implies that the eigenvectors of $\mathcal{L}_{\rm SA}$ are classified into 
those belonging to $\mathsf{P}$ and $\mathsf{Q}$. 
Equation~(\ref{PRP}) implies that 
$\mathcal{P} \mathcal{L}_{\rm SA} \mathcal{P} = \mathcal{P} \mathcal{L} \mathcal{P}$ holds, 
and that $\mathcal{R}$ maps operators in $\mathsf{P}$ to ones in $\mathsf{Q}$ 
($\mathcal{R} \mathcal{P} = \mathcal{Q} \mathcal{R} \mathcal{P}$).

\section{Perturbative Method for NESS}\label{section:main}

In the present paper, we are interested in the NESS $\hat{\rho}_{\rm ss}$ of the QME. 
By substituting $\hat{\rho}_{\rm ss}$ into the QME (\ref{QME}), 
we obtain the steady-state equation (since $\hat{\rho}_{\rm ss}$ is time-independent), 
\begin{align}
\mathcal{L} \hat{\rho}_{\rm ss} = 0.
\label{ssEquation}
\end{align}
Thus, our task is to solve the zero-eigenvalue problem of $\mathcal{L}$. 

For this purpose, we take the following strategy that consists of two steps.  
In the first step, we solve the zero-eigenvalue problem of $\mathcal{L}_{\rm SA}$, 
i.e., $\mathcal{L}_{\rm SA} \hat{\rho}_{\rm ss}^{\rm SA} = 0$. 
Here, we assume that there exists a unique $\hat{\rho}_{\rm ss}^{\rm SA}$. 
See Refs.~\cite{SpohnLebowitz,Spohn} and \cite{Evans} for the conditions for this assumption. 
We may solve this problem by either numerical or analytical methods. 
As mentioned earlier, the steady state $\hat{\rho}_{\rm ss}^{\rm SA}$ in the secular approximation 
is not suitable for analyzing NESS. 
However, combined with the following second step, $\hat{\rho}_{\rm ss}^{\rm SA}$ 
is a useful starting point to obtain NESS. 

In the second step, taking $\hat{\rho}_{\rm ss}^{\rm SA}$ as the unperturbed eigenvector, 
we perform the first-order perturbation calculation with respect to $v \mathcal{R}$ [see Eq.~(\ref{generator2})]. 
Then, we have an $O(v)$ correction to $\hat{\rho}_{\rm ss}^{\rm SA}$, 
which gives a perturbative solution of Eq.~(\ref{ssEquation}). 
This perturbation theory is valid up to the first order in $v$, which is the same order as that in the QME (\ref{QME}). 

This strategy shares the spirit in Ref.~\cite{Sugita} 
of solving the steady-state equation [Eq.~(\ref{ssEquation})] by a perturbative method that is valid up to $O(v)$. 
The difference lies in the decomposition into unperturbed and perturbation parts.

\subsection{Perturbation theory with respect to $v \mathcal{R}$}

Now, we analyze in detail the perturbation theory mentioned in the second step above. 
First, we define the eigenvalues $\lambda_k$ and the corresponding left and right eigenvectors 
$\hat{\ell}_k$ and $\hat{r}_k$ of the unperturbed generator $\mathcal{L}_{\rm SA}$ as  
\begin{align}
\mathcal{L}_{\rm SA}^\dag \hat{\ell}_k &= \overline{\lambda_k} \hat{\ell}_k, 
\\
\mathcal{L}_{\rm SA} \hat{r}_k &= \lambda_k \hat{r}_k. 
\end{align}
We assign $k=0$ to the zero eigenvalue; 
i.e., $\lambda_0=0$, $\hat{\ell}_0=\hat{1}$ (identity operator on $\mathsf{H}$), 
and $\hat{r}_0=\hat{\rho}_{\rm ss}^{\rm SA}$ is the steady solution of $\mathcal{L}_{\rm SA}$. 

Next, we apply a formal perturbation theory to the zero eigenvalue and eigenvectors of  $\mathcal{L}_{\rm SA}$ 
with respect to $v \mathcal{R}$. 
Note that $\lambda_0$ is nondegenerate because we assume that $\hat{r}_0$ is uniquely determined. 
Therefore, similarly to the case of quantum mechanics \cite{Messiah}, 
we can use the perturbation theory for the nondegenerate case 
to obtain the first-order terms $\delta \lambda$, $\widehat{\delta \ell}$, and $\widehat{\delta r}$ 
(corrections to $\lambda_0$, $\hat{\ell}_0$, and $\hat{r}_0$, respectively) as 
\begin{align}
\delta \lambda &= v {\rm Tr_S} \bigl[ \hat{\ell}_0^\dag \mathcal{R} \hat{r}_0 \bigr], 
\label{eigenValue0}
\\
\widehat{\delta\ell} 
&= v \sum_{k\neq 0} \overline{ \left( \frac{{\rm Tr_S} [\hat{\ell}_0^\dag \mathcal{R} \hat{r}_k ]}
{\lambda_0 - \lambda_k} \right) } \hat{\ell}_k, 
\label{leftEigen0}
\\
\widehat{\delta r} 
&= v \sum_{k\neq 0} \frac{{\rm Tr_S} [\hat{\ell}_k^\dag \mathcal{R} \hat{r}_0 ]}
{\lambda_0 - \lambda_k} \hat{r}_k. 
\label{rightEigen0}
\end{align}
Since $\hat{\ell}_0=\hat{1}$, Eq.~(\ref{eigenValue0}) and the numerators in Eq.~(\ref{leftEigen0}) have the form of 
$v{\rm Tr_S} \bigl[ \mathcal{R} \hat{A} \bigr]$. 
By using $v\mathcal{R} = \mathcal{L} - \mathcal{L}_{\rm SA}$ and 
the trace-preserving property of the Redfield and Lindblad QMEs, 
we have $v{\rm Tr_S} \bigl[ \mathcal{R} \hat{A} \bigr]=0$.
Therefore, the correction terms $\delta\lambda$ and $\widehat{\delta\ell}$ vanish. 
This fact indicates that the eigenvalue and left eigenvector remain respectively zero and $\hat{1}$ 
in this perturbation theory, 
and thus the corresponding right eigenvector $\hat{r}_0 + \widehat{\delta r}$ 
is indeed the steady state of $\mathcal{L}$. 

Next, we investigate the correction term $\widehat{\delta r}$ to the right eigenvector. 
As mentioned earlier concerning Eq.~(\ref{SA-L}), 
each eigenvector of $\mathcal{L}_{\rm SA}$ belongs to either $\mathsf{P}$ or $\mathsf{Q}$. 
In particular, $\hat{r}_0$ belongs to $\mathsf{P}$ since ${\rm Tr_S}\hat{r}_0=1$ 
($\hat{r}_0=\hat{\rho}_{\rm ss}^{\rm SA}$ is a density matrix). 
This fact with Eq.~(\ref{PRP}) leads to $\mathcal{R} \hat{r}_0 \in \mathsf{Q}$ 
($\mathcal{R} \hat{r}_0 = \mathcal{R} \mathcal{P} \hat{r}_0 = \mathcal{Q} \mathcal{R} \mathcal{P} \hat{r}_0$). 
Therefore, the terms with $\hat{\ell}_k \in \mathsf{P}$ in Eq.~(\ref{rightEigen0}) vanish. 
Thus, we obtain 
\begin{align}
\widehat{\delta r} = - v \sum_{k\in \mathsf{Q}} 
\frac{{\rm Tr_S} [\hat{\ell}_k^\dag \mathcal{R} \hat{\rho}_{\rm ss}^{\rm SA} ]}{\lambda_k} \hat{r}_k, 
\end{align}
where the sum runs over the labels whose eigenvectors belong to $\mathsf{Q}$. 
To rewrite $\widehat{\delta r}$ further, we evaluate $\lambda_k$, $\hat{\ell}_k$, and $\hat{r}_k$ in the above equation. 
To this end, we note that it is sufficient to evaluate them to $O(v^0)$ 
because $\widehat{\delta r}$ includes $v$ in the prefactor. 
Therefore, we can replace $\lambda_k$, $\hat{\ell}_k$, and $\hat{r}_k$ 
with the eigenvalue and eigenvectors of $\mathcal{L}_0$. 
Since the eigenvalue is $(E_i-E_j)/i\hbar$ 
and both the left and right eigenvectors are $|E_i, m_i\rangle \langle E_j, m_j|$, 
we finally obtain 
\begin{align}
\widehat{\delta r} 
&= - i\hbar v \sum_{i \neq j} \sum_{m_i, m_j} 
\frac{\langle E_i, m_i| \mathcal{R} \hat{\rho}_{\rm ss}^{\rm SA} |E_j, m_j\rangle }{E_i-E_j} |E_i, m_i\rangle \langle E_j, m_j|. 
\label{1stCorrection}
\end{align}
This result implies two important points. 
One is that all the denominators in Eq.~(\ref{1stCorrection}) are $O(v^0)$. 
This is necessary for the validity of this perturbation theory 
since if one of the denominators is on the order of $v$, $\widehat{\delta r}$ becomes on the order of $v^0$, 
which is contradictory to the assumption that this is the first-order term. 
The other point is that, to calculate $\widehat{\delta r}$, 
we have to determine not all the eigenvalues and eigenvectors of $\mathcal{L}_{\rm SA}$ 
but only the right eigenvector for the zero eigenvalue.

We thus construct a perturbative method of calculating the NESS $\hat{\rho}_{\rm ss}$ of the QME (\ref{QME}): 
\begin{align}
\hat{\rho}_{\rm ss} = \hat{\rho}_{\rm ss}^{\rm SA} + \widehat{\delta r} 
\label{mainResult}
\end{align} 
(or equivalently, $\mathcal{P} \hat{\rho}_{\rm ss} = \hat{\rho}_{\rm ss}^{\rm SA}$
and $\mathcal{Q} \hat{\rho}_{\rm ss} = \widehat{\delta r}$) with Eq.~(\ref{1stCorrection}). 
This is the main result in the present paper. 
We note that this result is very similar to the result in Ref.~\cite{Thingna_etal}.

\subsection{Advantage in numerical computation}

The method of Eq.~(\ref{mainResult}) requires information on $\hat{\rho}_{\rm ss}^{\rm SA}$, 
which is the right eigenvector of $\mathcal{P} \mathcal{L}_{\rm SA} \mathcal{P}$ 
($=\mathcal{P} \mathcal{L} \mathcal{P}$) for the zero eigenvalue. 
Therefore, when we use this method in calculating the NESS $\hat{\rho}_{\rm ss}$ of the QME (\ref{QME}), 
we need not directly solve the full steady-state equation (zero-eigenvalue problem) of $\mathcal{L}$, 
but we have to solve the reduced zero-eigenvalue problem of $\mathcal{P} \mathcal{L} \mathcal{P}$. 
Since the dimension of the matrix $\mathcal{P} \mathcal{L} \mathcal{P}$ ($\dim\mathsf{P}$) is much smaller than 
that of $\mathcal{L}$ ($\dim\mathsf{L}$), 
this method has an advantage in the numerical computation of $\hat{\rho}_{\rm ss}$. 

The computation of $\widehat{\delta r}$ with Eq.~(\ref{1stCorrection}) requires information on 
$\mathcal{R} \hat{\rho}_{\rm ss}^{\rm SA} = \mathcal{Q} \mathcal{R} \mathcal{P} \hat{\rho}_{\rm ss}^{\rm SA}$. 
Although the matrix $\mathcal{Q} \mathcal{R} \mathcal{P}$ ($\dim\mathsf{Q} \times \dim\mathsf{P}$) 
is not significantly smaller than $\mathcal{L}$, 
the computational cost of the matrix-vector multiplication is much smaller than that of solving the eigenvalue problem. 

In Ref.~\cite{Wichterich_etal}, Wichterich {\it et al}. proposed a Lindblad-type QME 
that has nonvanishing internal current in NESS, 
where they assumed weak internal couplings and high temperature. 
In contrast, our method does not require these assumptions. 
When using a stochastic unraveling method \cite{BreuerPetruccione,Carmichael,PlenioKnight}, 
one can reduce the problem size from $\dim\mathsf{L}$ to $\dim\mathsf{H}$. 
However, it requires a Monte Carlo calculation; i.e., one must take the average of many runs of computation. 
By comparison, our method determines the NESS in a single run.

\subsection{Current}

\begin{figure}[t]
\begin{center}
\includegraphics[width=0.6\linewidth]{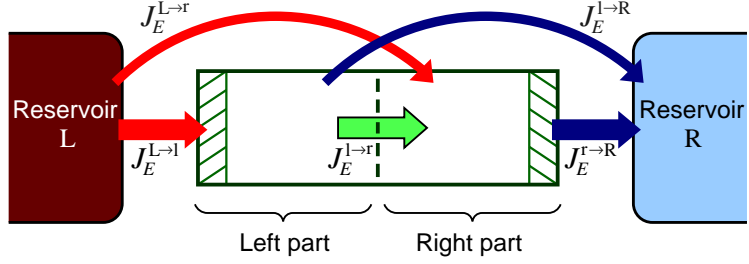}
\end{center}
\caption{
(Color online) Currents in the open system setup.
The dashed line represents a virtual division of the system S into the left and right parts. 
In the main text, the operator identity $\hat{J}_E^{\rm L \to r} = \hat{J}_E^{\rm l \to R} = 0$
and the steady-state average equality $\langle \hat{J}_E^{\rm l \to r} \rangle_{\rm ss} 
= \langle \hat{J}_E^{\rm L \to l}\rangle_{\rm ss} = \langle \hat{J}_E^{\rm r \to R} \rangle_{\rm ss}$
are shown.
}
\label{fig:current}
\end{figure}

Here, we consider energy currents in the present formulation. 
As shown in Fig.~\ref{fig:current}, we virtually divide the system S into two parts; 
the left part includes the interaction region with the reservoir L, 
while the right part includes the interaction region with the reservoir R. 
Correspondingly to this division, we can write the system Hamiltonian as 
$\hat{H}_{\rm S} = \hat{H}_{\rm l} + \hat{H}_{\rm r}$, 
where $\hat{H}_{\rm l}$ and $\hat{H}_{\rm r}$ are respectively the Hamiltonian of the left and right parts. 
The interaction between the left and right parts can be included
in either $\hat{H}_{\rm l}$ or $\hat{H}_{\rm r}$.
We then define the energy current $\hat{J}_E^{\rm l \to r}$ from the left part to the right part as the unit-time loss of 
the left energy $\hat{H}_{\rm l}$ due to the interaction with the right part: 
$\hat{J}_E^{\rm l \to r} \equiv [\hat{H}_{\rm r} , \hat{H}_{\rm l}]/i\hbar = [\hat{H}_{\rm S} , \hat{H}_{\rm l}]/i\hbar 
= - \mathcal{L}_0^\dag \hat{H}_{\rm l}$. 
This is equivalent to the unit-time gain of the right energy $\hat{H}_{\rm r}$ 
due to the interaction with the left part, 
$\hat{J}_E^{\rm l \to r} = - [\hat{H}_{\rm l} , \hat{H}_{\rm r}]/i\hbar = \mathcal{L}_0^\dag \hat{H}_{\rm r}$, 
because the energy conservation holds if the interactions with the reservoirs are absent 
($\mathcal{L}_0^\dag \hat{H}_{\rm S} = 0$). 

We note that ${\rm Tr} [ \hat{J}_E^{\rm l \to r} \hat{\rho}_{\rm ss}^{\rm SA} ]=0$ holds 
(because $\hat{J}_E^{\rm l \to r} \in \mathsf{Q}$ and  $\hat{\rho}_{\rm ss}^{\rm SA} \in \mathsf{P}$), 
which indicates that the internal current vanishes in the SA-QME, as mentioned earlier. 
Using this fact, we have the steady-state average of the current in the Redfield QME as 
\begin{align}
\langle \hat{J}_E^{\rm l \to r} \rangle_{\rm ss} 
&= {\rm Tr} \bigl[ \hat{J}_E^{\rm l \to r} \hat{\rho}_{\rm ss} \bigr] 
=  {\rm Tr} \bigl[ \hat{J}_E^{\rm l \to r} \widehat{\delta r} \bigr] 
\nonumber\\
&= v \sum_{i \neq j} \sum_{m_i, m_j} \langle E_j, m_j| \hat{H}_{\rm l} |E_i, m_i\rangle 
\langle E_i, m_i| \mathcal{R} \hat{\rho}_{\rm ss}^{\rm SA} |E_j, m_j\rangle 
\nonumber\\
&= v {\rm Tr} \bigl[ \hat{H}_{\rm l} \mathcal{R} \hat{\rho}_{\rm ss}^{\rm SA} \bigr].
\end{align}
Here, we used  Eq.~(\ref{1stCorrection}) for $\widehat{\delta r}$ in the second line 
and $\mathcal{R} \hat{\rho}_{\rm ss}^{\rm SA} \in \mathsf{Q}$ in the third line. 

We next investigate the energy current from the reservoir L to the system S. 
We define the current operator as the unit-time energy gain of S 
due to the coupling with the reservoir L: 
$\hat{J}_E^{\rm L \to S} \equiv v \mathcal{L}_{\rm L}^\dag \hat{H}_{\rm S}$. 
We can divide it as $\hat{J}_E^{\rm L \to S} = \hat{J}_E^{\rm L \to l} + \hat{J}_E^{\rm L \to r}$, 
where $\hat{J}_E^{\rm L \to l} \equiv v \mathcal{L}_{\rm L}^\dag \hat{H}_{\rm l}$ 
is the current from the reservoir L to the left part of S, 
and $\hat{J}_E^{\rm L \to r} \equiv v \mathcal{L}_{\rm L}^\dag \hat{H}_{\rm r}$ 
is the current from the reservoir L to the right part (see Fig.~\ref{fig:current}). 
However, $\hat{J}_E^{\rm L \to r}=0$ holds 
if the right part of S is sufficiently separate from the interaction region with the reservoir L 
(the hatched region in the left part in Fig.~\ref{fig:current}). 
We note that this condition is usually satisfied (except for extremely small systems). 
We can show the operator identity $\hat{J}_E^{\rm L \to r}=0$ as 
\begin{align}
\hat{J}_E^{\rm L \to r} 
&= \frac{v}{2\hbar^2} \sum_{\lambda,\nu} \sum_\omega \Xi_{\lambda\nu}^{\rm L} (\omega) 
\Bigl\{ 
\hat{X}_{{\rm L},\lambda} \hat{H}_{\rm r} \hat{X}_{{\rm L},\nu}^{(\omega)\dag} 
- \hat{H}_{\rm r} \hat{X}_{{\rm L},\lambda} \hat{X}_{{\rm L},\nu}^{(\omega)\dag} 
\Bigr\} + {\rm h.c.}
\nonumber\\ 
&=0. 
\end{align} 
Here, we used the commutability of $\hat{H}_{\rm r}$ with $\hat{X}_{{\rm L},\lambda}$ 
(because of the lack of overlap between the right part and the interaction region with the reservoir L). 

Similarly, we can define the energy current from the system S to the reservoir R 
$\hat{J}_E^{\rm S \to R} \equiv -v \mathcal{L}_{\rm R}^\dag \hat{H}_{\rm S}$, 
and show that it is equivalent to the current from the right part to the reservoir R 
$\hat{J}_E^{\rm r \to R} \equiv -v \mathcal{L}_{\rm R}^\dag \hat{H}_{\rm r}$ 
because the current from the left part to the reservoir R is absent 
($\hat{J}_E^{\rm l \to R} \equiv -v \mathcal{L}_{\rm R}^\dag \hat{H}_{\rm l}=0$). 

By using these facts, we can show the continuity equations of the energy in the Heisenberg picture: 
$(d/dt) \hat{H}_{\rm l} = -\hat{J}_E^{\rm l \to r} + \hat{J}_E^{\rm L \to l}$ 
and $(d/dt) \hat{H}_{\rm r} = \hat{J}_E^{\rm l \to r} - \hat{J}_E^{\rm r \to R}$. 
In the steady state,  we thus obtain 
$\langle \hat{J}_E^{\rm l \to r} \rangle_{\rm ss} = \langle \hat{J}_E^{\rm L \to l}\rangle_{\rm ss} 
= \langle \hat{J}_E^{\rm r \to R} \rangle_{\rm ss}$ as expected.

\section{Examples}\label{section:example}

\subsection{Analytical expression of NESS in noninteracting fermion system}

We consider a noninteracting fermion system that is connected with two reservoirs, as an example. 
The reservoirs are in equilibrium states at different temperatures and chemical potentials, 
so that the particle and energy currents flow in the system. 
Using the method of Eqs.~(\ref{mainResult}) and (\ref{1stCorrection}), 
we derive an analytical expression of the density matrix of the NESS. 

We consider a system S of fermions that move on a one-dimensional $N$-site lattice:  
\begin{align}
\hat{H}_{\rm S} = \sum_{l=1}^N \varepsilon_l \hat{d}_l^\dag \hat{d}_l 
+ \sum_{l=1}^{N-1} (t_l^{\rm hop} \hat{d}_l^\dag \hat{d}_{l+1} + {\rm h.c.}), 
\end{align}
where $\varepsilon_l$ is the energy level of the $l$th site, 
$t_l^{\rm hop}$ is the transfer probability amplitude between the $l$th and $(l+1)$th sites, 
and $\hat{d}_l^\dag$ and $\hat{d}_l$ are respectively the creation and annihilation operators 
of the fermion at the $l$th site. 
We can diagonalize this Hamiltonian by an appropriate linear transformation:
\begin{align}
\hat{H}_{\rm S} = \sum_{k=1}^N \hbar\omega_k \hat{c}_k^\dag \hat{c}_k, 
\end{align}
where
\begin{align}
\hat{d}_l = \sum_{k=1}^N W_{lk} \hat{c}_k, 
~~
\hat{d}_l^\dag = \sum_{k=1}^N \overline{W}_{lk} \hat{c}_k^\dag, 
\label{transformedOp}
\end{align}
and $W_{lk}$ is the transformation matrix. 
For simplicity, we assume that $\omega_k \neq \omega_{k'}$ if $k \neq k'$. 
In this case, $\hat{H}_{\rm S}$ is nondegenerate. 

At the left ($l=1$) and right ($l=N$) ends of the lattice, the system S is respectively coupled 
to the reservoirs L (at the inverse temperature $\beta_{\rm L}$ and the chemical potential $\mu_{\rm L}$) 
and R (at the inverse temperature $\beta_{\rm R}$ and the chemical potential $\mu_{\rm R}$). 
We assume that these reservoirs are free fermion systems:
\begin{align}
\hat{H}_b = \sum_q \varepsilon_{b,q} \hat{c}_{b,q}^\dag \hat{c}_{b,q}, 
\end{align}
where $\hat{c}_{b,q}^\dag$ and $\hat{c}_{b,q}$ are the creation and annihilation fermion operators 
in the reservoir $b$ $(b={\rm L,R})$, respectively. 
The coupling Hamiltonian is given by 
\begin{align}
\hat{H}_{{\rm S}b} &= \sum_q  \bigl( \hbar \xi_{b,q} \hat{d}_{l(b)}^\dag \hat{c}_{b,q} + {\rm h.c.} \bigr) 
\nonumber\\
&= \sum_{k,q} \bigl( \hbar \zeta_{kbq} \hat{c}_k^\dag \hat{c}_{b,q} + {\rm h.c.} \bigr). 
\end{align}
In the second line, we used Eq.~(\ref{transformedOp}).
Here, $\xi_{b,q}$ is the coupling constant, $\zeta_{kbq} \equiv \xi_{b,q} \overline{W}_{l(b)k}$, 
$l(b)=1$ if $b={\rm L}$, and $l(b)=N$ if $b={\rm R}$. 
We note that, unlike Eq.~(\ref{interaction}), $\hat{c}_k^\dag$ is not self-adjoint. 
However, the generic formulation in Sects.~\ref{section:setup} and \ref{section:main} is unmodified by this difference. 
In this case, it is convenient to use a slightly different definition of eigenoperators: 
\begin{align}
\hat{c}_k^{(\omega)} &\equiv \sum_i \hat{\Pi}(E_i) \hat{c}_k \hat{\Pi}(E_i+\hbar\omega), 
\nonumber\\
\hat{c}_k^{\dag(\omega)} &\equiv \sum_i \hat{\Pi}(E_i) \hat{c}_k^\dag \hat{\Pi}(E_i-\hbar\omega). 
\end{align}
In this definition, $(\hat{c}_k^{(\omega)})^\dag = \hat{c}_k^{\dag(\omega)}$ holds.
In particular, in the case of the noninteracting fermion system, 
$\hat{c}_k^{(\omega)} = \hat{c}_k \delta_{\omega,\omega_k}$ 
and $\hat{c}_k^{\dag(\omega)} = \hat{c}_k^\dag \delta_{\omega,\omega_k}$. 
In almost the same manner as in Sect.~\ref{section:setup}, we obtain the Redfield QME generator $\mathcal{L}$, 
where $\mathcal{L} = \mathcal{L}_0 + v \sum_b \mathcal{L}_b$ 
with $\mathcal{L}_0 \hat{\rho} = [\hat{H}_{\rm S}, \hat{\rho}]/i\hbar$ and 
\begin{align}
\mathcal{L}_b \hat{\rho} = \frac{1}{2} \sum_{k_1,k_2} \sum_\omega 
\Gamma_b(\omega) 
& \Bigl\{ \overline{W}_{l(b)k_1} W_{l(b)k_2} f^+_b(\omega) 
\bigl( \hat{c}_{k_1}^{\dag(\omega)} \hat{\rho} \hat{c}_{k_2} + \hat{c}_{k_1}^\dag \hat{\rho} \hat{c}_{k_2}^{(\omega)} 
- \hat{c}_{k_2} \hat{c}_{k_1}^{\dag(\omega)} \hat{\rho} - \hat{\rho} \hat{c}_{k_2}^{(\omega)} \hat{c}_{k_1}^\dag \bigr)
\nonumber\\
&+ W_{l(b)k_1} \overline{W}_{l(b)k_2} f^-_b(\omega) 
\bigl( \hat{c}_{k_1}^{(\omega)} \hat{\rho} \hat{c}_{k_2}^\dag + \hat{c}_{k_1} \hat{\rho} \hat{c}_{k_2}^{\dag(\omega)} 
- \hat{c}_{k_2}^\dag \hat{c}_{k_1}^{(\omega)} \hat{\rho} - \hat{\rho} \hat{c}_{k_2}^{\dag(\omega)} \hat{c}_{k_1} \bigr) 
\Bigr\}. 
\nonumber
\end{align}
Here, $\Gamma_b(\omega) \equiv 2\pi \sum_q |\xi_{b,q}|^2 \delta(\omega-\varepsilon_{b,q}/\hbar)$ 
is the reservoir spectral function and $f^\pm_b(\omega) \equiv 1/(1+e^{\pm\beta_b(\hbar\omega-\mu_b)})$. 
In the above equation, we omitted terms proportional to the imaginary part of $\Xi$ \cite{Saito_etal_2}, 
since they do not have a significant contribution. 
Also, the SA-QME generator $\mathcal{L}_{\rm SA}$ is given as 
$\mathcal{L}_{\rm SA} = \mathcal{L}_0 + v \sum_b \mathcal{L}_b^{\rm SA}$, 
where 
\begin{align}
\mathcal{L}_b^{\rm SA} \hat{\rho} = \frac{1}{2} \sum_k 
\gamma_{kb} \Bigl\{ f^+_b(\omega_k) \bigl( 2 \hat{c}_k^\dag \hat{\rho} \hat{c}_k 
- \hat{c}_k \hat{c}_k^\dag \hat{\rho} - \hat{\rho} \hat{c}_k \hat{c}_k^\dag \bigr)
+  f^-_b(\omega_k) \bigl( 2 \hat{c}_k \hat{\rho} \hat{c}_k^\dag 
- \hat{c}_k^\dag \hat{c}_k \hat{\rho} - \hat{\rho} \hat{c}_k^\dag \hat{c}_k \bigr) \Bigr\}, 
\nonumber
\end{align}
with $\gamma_{kb} = \Gamma_b(\omega_k) \big| W_{l(b)k} \big|^2$. 

We can solve the zero-eigenvalue problem of the above $\mathcal{L}_{\rm SA}$ 
with the corresponding eigenvector of the form of $\hat{\rho}_{\rm ss}^{\rm SA} = \bigotimes_k \hat{\rho}_k$, where 
\begin{align}
\hat{\rho}_k &= \frac{1}{\sum_b \gamma_{kb}} 
\sum_b \gamma_{kb} \Bigl[ f_b^-(\omega_k) \hat{c}_k \hat{c}_k^\dag 
+ f_b^+(\omega_k) \hat{c}_k^\dag \hat{c}_k \Bigr]. 
\end{align}
This is equivalent to a weighted average of the equilibrium states 
with the parameters corresponding to the reservoir L or R: 
$\hat{\rho}_k = \sum_b \gamma_{kb} \hat{\rho}_{kb}^{\rm eq} / \sum_b \gamma_{kb}$. 
(Note that the equilibrium state for this model at the inverse temperature $\beta_b$ 
and the chemical potential $\mu_b$ is written as 
$\hat{\rho}_b^{\rm eq} = \exp[-\beta_b \sum_k (\hbar\omega_k-\mu_b) \hat{c}_k^\dag \hat{c}_k]/Z_b^{\rm eq} 
= \bigotimes_k [f_b^-(\omega_k) \hat{c}_k \hat{c}_k^\dag + f_b^+(\omega_k) \hat{c}_k^\dag \hat{c}_k] 
\equiv \bigotimes_k \hat{\rho}_{kb}^{\rm eq}$). 
We can also express $\hat{\rho}_{\rm ss}^{\rm SA}$ as 
\begin{align}
\hat{\rho}_{\rm ss}^{\rm SA} &= \frac{1}{Z_{\rm ss}} \exp \Bigl[ -\sum_k G_k \hat{c}_k^\dag \hat{c}_k \Bigr], 
\label{rho_ss_SA_freeFermion}
\end{align}
where $G_k \equiv \ln \sum_b \gamma_{kb} f_b^-(\omega_k) - \ln \sum_b \gamma_{kb} f_b^+(\omega_k)$. 
This result is equivalent to the result in Sect.~III A of Ref.~\cite{Dhar_etal}.

By using Eq.~(\ref{1stCorrection}) with this result, we obtain the first-order correction of the NESS density matrix: 
\begin{align}
\widehat{\delta r} 
= & v \frac{i}{2} \sum_{k_1 \neq k_2} \frac{1}{\omega_{k_1} - \omega_{k_2}} 
\biggl[ 
\Bigl\{ F_{k_1 k_2}^+ \hat{c}_{k_1} \hat{c}_{k_2}^\dag \hat{\rho}_{\rm ss}^{\rm SA} 
- F_{k_1 k_2}^- \hat{\rho}_{\rm ss}^{\rm SA} \hat{c}_{k_1} \hat{c}_{k_2}^\dag \Bigr\} 
- {\rm h.c.}
\biggr], 
\label{rho_ss_1stC_freeFermion}
\end{align}
where 
\begin{align}
F_{k_1 k_2}^\pm \equiv \frac{\sum_b \gamma_{k_2b}}{\sum_b \gamma_{k_2b} f^\pm_b(\omega_{k_2})}
\sum_b \overline{W}_{l(b)k_1} W_{l(b)k_2} \Gamma_b(\omega_{k_1}) f^\pm_b(\omega_{k_1}). 
\nonumber
\end{align}
Equations~(\ref{rho_ss_SA_freeFermion}) and (\ref{rho_ss_1stC_freeFermion}) are respectively 
analytical expressions of the diagonal and off-diagonal elements of the NESS density matrix 
in the noninteracting fermion system.

\subsection{Numerical calculation in quantum spin chain}

As another example, we apply our method 
to the numerical computation of the NESS in a one-dimensional Ising spin chain subject to a magnetic field. 
The Hamiltonian of the chain system S reads 
\begin{align}
\hat{H}_{\rm S} = - J \sum_{l=1}^{N-1} \hat{\sigma}_l^z \hat{\sigma}_{l+1}^z 
- h_x \sum_{l=1}^N \hat{\sigma}_l^x - h_z \sum_{l=1}^N \hat{\sigma}_l^z, 
\label{tiltedFieldIsingModel}
\end{align}
where $\hat{\sigma}_l^\alpha$ ($\alpha = x,z$) is the $\alpha$-component of the Pauli matrices at the $l$th site, 
$J$ is the Ising coupling constant, $h_\alpha$ is the $\alpha$-component of the magnetic field, 
and $N$ is the total site number in the chain. 
At the left ($l=1$) and right ($l=N$) ends of the chain, 
the system S is coupled to the heat reservoirs L (at the inverse temperature $\beta_{\rm L}$) 
and R (at the inverse temperature $\beta_{\rm R}$), respectively. 
We assume that these reservoirs are composed of  free boson particles: 
\begin{align}
\hat{H}_b = \sum_k \varepsilon_{b,k} \hat{a}_{b,k}^\dag \hat{a}_{b,k},
\end{align}
where $\hat{a}_{b,k}^\dag$ and $\hat{a}_{b,k}$ are the creation and annihilation boson operators in the reservoir $b$ 
($b={\rm L,R}$), respectively. 
The system-reservoir coupling Hamiltonian reads 
\begin{align}
\hat{H}_{{\rm S}b} = \hat{\sigma}^x_{l(b)} \sum_k \hbar (\xi_{b,k} \hat{a}_{b,k}^\dag + \xi_{b,k}^* \hat{a}_{b,k}), 
\end{align}
where $\xi_{b,k}$ is the coupling constant, $l(b)=1$ if $b={\rm L}$, and $l(b)=N$ if $b={\rm R}$. 

Then, from Eqs.~(\ref{generator1}) and (\ref{dissipator2}), the generator of the Redfield QME reads 
$\mathcal{L} = \mathcal{L}_0 + v \sum_b \mathcal{L}_b$ 
with $\mathcal{L}_0 \hat{\rho} = [\hat{H}_{\rm S}, \hat{\rho}]/i\hbar$ and 
\begin{align}
\mathcal{L}_b \hat{\rho} 
= \frac{1}{2} \sum_\omega \Phi_b(\omega) 
\Bigl\{ 
\hat{\sigma}_{l(b)}^x(-\omega) \hat{\rho} \hat{\sigma}_{l(b)}^x 
+ \hat{\sigma}_{l(b)}^x \hat{\rho} \hat{\sigma}_{l(b)}^x(\omega) 
- \hat{\sigma}_{l(b)}^x \hat{\sigma}_{l(b)}^x(-\omega) \hat{\rho} 
- \hat{\rho} \hat{\sigma}_{l(b)}^x(\omega) \hat{\sigma}_{l(b)}^x
\Bigr\}. 
\nonumber
\end{align}
Here, $\hat{\sigma}_{l(b)}^x(\omega)=\sum_i \hat{\Pi}(E_i) \hat{\sigma}_{l(b)}^x \hat{\Pi}(E_i + \hbar\omega)$ 
is the eigenoperator of $\hat{H}_{\rm S}$. 
The Fourier transform of the reservoir correlation function is given by 
$\Phi_b(\omega) = \bigl[ \Gamma_b(\omega) - \Gamma_b(-\omega) \bigr] n_b(\omega)$, 
where $\Gamma_b(\omega) = 2\pi \sum_k |\xi_{b,k}|^2 \delta(\omega-\varepsilon_{b,k}/\hbar)$ 
is the spectral function of the reservoir 
and $n_b (\omega) = 1/(e^{\beta_b\hbar\omega}-1)$ is the Bose distribution function. 
In the above equation, we omitted terms proportional to the imaginary part of $\Xi$ \cite{Saito_etal_2}, 
since they do not have a significant contribution. 

\begin{figure}[t]
\begin{center}
\includegraphics[width=0.6\linewidth]{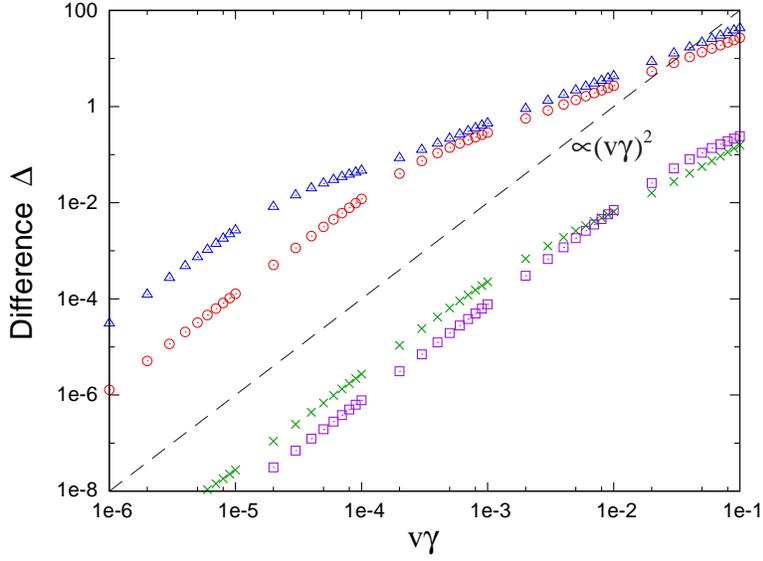}
\end{center}
\caption{
(Color online) Difference $\Delta$ between the NESSs calculated with and without the perturbative method 
[see Eq.~(\ref{deviation})]. 
In the numerical computation, $h_x$, $\hbar$, and $k_{\rm B}$ are set to unity. 
The system size is $N=6$. 
The reservoir temperatures are $1/\beta_{\rm L} = 2$ and $1/\beta_{\rm R} = 0.5$. 
The other parameters ($J, h_z$) are set to (0.1, 1.0) for triangles, 
(1.0, 1.0) for circles, (1.0, 0.1) for crosses, and (0.1, 0.1) for squares. 
The dashed line is an eye guide that is proportional to $(v\gamma)^2$. 
}
\label{fig:Ising}
\end{figure}

We numerically compute the NESS $\hat{\rho}_{\rm ss}^{\rm per}$ 
using Eqs.~(\ref{mainResult}) and (\ref{1stCorrection}) 
and compare it with the NESS $\hat{\rho}_{\rm ss}^{\rm ori}$ 
that we obtain by directly solving the steady-state equation of $\mathcal{L}$. 
In the computations, we use the Ohmic spectral function 
$\Gamma_b(\omega)=\gamma \omega \Theta(\omega)$, where $\Theta(\omega)$ is the step function. 
We set the system size to $N=6$ and the temperatures to $1/\beta_{\rm L} = 2$ and $1/\beta_{\rm R} = 0.5$
(in the units of $h_x=1$, $\hbar=1$, and $k_{\rm B}=1$). 
In Fig.~\ref{fig:Ising}, we plot the sum of the differences in the density matrix elements 
\begin{align}
\Delta = \sum_{i,j} 
\Big| \langle E_i | \hat{\rho}_{\rm ss}^{\rm per} |E_j \rangle 
- \langle E_i | \hat{\rho}_{\rm ss}^{\rm ori} |E_j \rangle \Big| 
\label{deviation}
\end{align}
as a function of $v\gamma$ 
($\hat{H}_{\rm S}$ does not have degeneracy for the parameters used in the computations). 
We observe that, in the small $v\gamma$ region, the difference $\Delta$ decreases in proportion to $(v\gamma)^2$ 
as $v\gamma$ becomes smaller. 
Therefore, as expected, we conclude that $\hat{\rho}_{\rm ss}^{\rm per}$ 
is in agreement with $\hat{\rho}_{\rm ss}^{\rm ori}$ within the error on the order of $(v\gamma)^2$.

\section{Concluding Remarks}\label{section:conclusion}

In the present study, we have formulated a method of computing NESS of the Redfield QME (\ref{QME}). 
It is a perturbation theory in $v$ ($=u^2$), the order of which is the same as that of the QME (first order in $v$). 

We have used the method to derive an analytical expression of the NESS density matrix 
in a noninteracting fermion system connected to two fermion reservoirs. 
We have also applied the method to the numerical computation of the NESS 
in a spin chain connected to two heat reservoirs at different temperatures. 
With the numerical computation, we have demonstrated the validity of the method. 

Although we have formulated the above method in a setup of a system coupled to two reservoirs, 
it is straightforward to extend it to the cases with more reservoirs. 
We have also discussed energy currents in this formulation. 
We can extend the discussion to other cases, such as particle and spin currents. 
We also note that the present method for NESS is applicable even to non-Markovian cases, 
although we have formulated it within a Markovian QME. 
This is because the Markovian approximation is good when the system changes slowly \cite{KuboTodaHashitsume}, 
and hence has no effect on the steady-state solution; 
therefore, up to first order in $v$, the steady state in the non-Markovian QME 
is equivalent to that in the Markovian QME. 

The system size available in the numerical computation by the present method is about twice larger 
than that by the methods of directly solving the zero-eigenvalue problem of the QME generator $\mathcal{L}$. 
The numerical limitation of the system size in the present method comes from 
the maximum matrix size of the system Hamiltonian $\hat{H}_{\rm S}$. 
Since the method requires all the eigenenergies and eigenstates of $\hat{H}_{\rm S}$, 
one should use the Householder method to diagonalize $\hat{H}_{\rm S}$. 
For systems composed of $S=1/2$ spins, the maximum system size would be 16 sites 
within the current computational ability. 
One of the ways to treat larger systems is to take only a relevant part of the eigenstates of $\hat{H}_{\rm S}$; 
for example, in low-temperature cases, it is expected to be sufficient to use only low-energy eigenstates 
(computed by, for example, the Lanczos method). 
This expectation should be verified in further investigation.

\section*{Acknowledgments}
This work was supported by a JSPS Research Fellowship for Young Scientists (No. 24-1112), 
by KAKENHI No. 26287087, 
and by ImPACT Program of Council for Science, Technology and Innovation (Cabinet Office, Government of Japan).


\begin{thebibliography}{99}
\bibitem{note1}
For the current between the system S and reservoirs, 
it is shown that the expectation value and full counting statistics 
in the SA-QME are equivalent to those in the original Redfield QME \cite{Yuge_etal_2}.
\end{thebibliography}
\end{document}